# Automating Discovery of Dominance in Synchronous Computer-Mediated Communication

[ Preprint ]

Jim Samuel
Baruch College, CUNY, NY
jim.samuel@baruch.cuny.edu

Richard Holowczak
Baruch College, CUNY, NY
richard.holowczak@baruch.cuny.edu

Raquel Benbunan-Fich
Baruch College, CUNY, NY
rbfich@baruch.cuny.edu

Ilan Levine
Levine Solutions, NY
ilanlevine@gmail.com

## Abstract

*With the advent of electronic interaction, dominance (or the assertion of control over others) has acquired new dimensions. This study investigates the dynamics and characteristics of dominance in virtual interaction by analyzing electronic chat transcripts of groups solving a hidden profile task. We investigate computer-mediated communication behavior patterns that demonstrate dominance and identify a number of relevant variables. These indicators are calculated with automatic and manual coding of text transcripts. A comparison of both sets of variables indicates that automatic text analysis methods yield similar conclusions than manual coding. These findings are encouraging to advance research in text analysis methods in general, and in the study of virtual team dominance in particular.*

## 1. Introduction

Dominance is a highly influential trait observable in individuals, groups and organizations [3]. In social interaction, dominance is an important dimension [8][9], and has traditionally been associated with the use of force, aggression and tactics with the intent of establishing superiority [22]. Liska [19] distinguishes *individual dominance*, which typically comes from age, inheritance, kinship, title-based roles or other characteristics, from *social dominance* which results from the application of individual abilities and skills in the context of 'social interaction displays'. Dominance is a richer construct when studied on an interpersonal dimension, within a context that combines personal, situational, and relational factors [4]. in computer-mediated group interaction, a number of studies have reported that dominance can significantly influence participant satisfaction and group productivity [9] [8] [16] [26] [15] [13].

The present study is focused on dominance actions, tactics and processes, which are identified as behavioral manifestations in Computer-Mediated Communication (CMC). We define dominance expressions broadly as latent dimensions of behavior, personality and status. These dominance manifestations seek to assert influence or control over communications, beliefs, people, processes, resources, and to obtain benefits through the use of technologies that facilitate virtual interaction.

To the best of our knowledge, this is the first study that identifies specific dominance manifestations and analyzes them based on automatic and manual coding methods. Compared to traditional textual analysis, the study of computer-mediated "chat language" requires specific techniques to handle abbreviations, shortened spellings, non-verbal cues (i.e. emoticons). These particular characteristics of chat communication provide important information about the nature, content and tone of communication [13] and should not be overlooked.

The objective of this study is to investigate dominance by analyzing chat transcripts of group communication and to compare alternative models to identify dominance in computer-mediated communication behavior. Given the ease of collecting automatic textual descriptors vis-à-vis the costs and resources typically associated with manual coding methods, a comparison of both approaches would provide useful insights for advancing research in a number of fields.

To describe the study, this paper is organized as follows. The next section reviews the relevant literature on dominance in CMC, including theoretical perspectives, empirical studies and a review of measurement approaches. The following section outlines the research methodology as well as the specific coding techniques followed in this study. The fourth part describes the data analysis, which is followed by the results. The paper concludes with discussion, limitations, implications and future research directions.

## 2. Literature Review

There are three main schools of thought regarding dominance: a) Dominance as a vested dynamic, wherein the entity demonstrating dominance does so on the basis of position or some intrinsic characteristic such as inherited authority [20]. b) Dominance as a personality trait, wherein the entity demonstrating dominance does so because of certain inherent personality-based propensities such as an inordinate desire to win an argument or acquire a resource [20] [29] [4]. c) Dominance based on an 'interactionist' perspective, which is found in the manifestation of the dominance-submission dimension of human interaction. In this perspective, dominance is identified only if accompanied by some level of corresponding submission in human-to-human interaction [4].

While we acknowledge the influence of the personality view of dominance, our study is anchored in the interactionist perspective. Accordingly, it draws insights from dominance in-face-to-face and in-person communication behavior, which has been a topic of significant research interest in social psychology [19].

One of the key theories, which frames dominance characteristics and distinguishes 'dominance' from 'power' is the Dyadic Power Theory (DPT) by Rollins and Bahr [30]. This theory explains how the dyadic nature of power between two parties leads to use of dominance tactics in human interaction. Burgoon, Johnson and Koch [6] have shown the applicability of the dyadic principle in DPT to groups, particularly when two parties or two opposing points of view emerge in participant interaction within the group. Although closely related, it is important to distinguish dominance from closely related constructs such as power. While power indicates the ability to control, dominance indicates either an attempt to gain control or a state of control without necessarily possessing the ability to control [3] [4].

The original DPT [30] posited a linear relationship between degree of perception of power and the degree of dominance tactics. This relation was further qualified by Dunbar [4], who posited a curvilinear relationship based on the premise that perceptions of very high (or very low) power would result in a non-linear expression of dominance. The curvilinear model was empirically tested and supported by Dunbar, Bippus and Young [12], who also demonstrated that near-equal perceptions of power resulted in maximum expressions of dominance. In group decision-making, power is related to knowledge and availability of information. Therefore, a setting where no participant has more (or better) information than someone else on the team should produce a balanced distribution of power. This setting would allow for a maximum expression of dominance propensities based on personality traits and situational demands.

### 2.1. Dominance in Electronic Communication

The study of Group Decision Support Systems has been one of the most prolific areas of research in IS. While it initially sought to compare the merits of face-to-face with computer-mediated communication, it later evolved to investigate in more depth the nature electronic interaction. In recent years, the focus has been to develop a more complete understanding of the nature of virtual teams and how computer-mediated interaction and group processes influence outcomes. Of this rich area of research, we identified a set of studies that provide insights for investigating dominance in computer-mediated group processes.

The first set of studies by Huang and Wei [14] [15] addressed group processes and posited that computer support for group communication enhanced the effects of informational influence, and diminished the effects of normative influence. In this context, they argued that dominance influence was mitigated through the use of an electronic medium. These studies provide evidence than an electronic medium acts as equalizer and likely contributes to reduce traditional dominance characteristics stemming from personal traits. However, no specific expressions of dominance are identified in these articles.

The second study by Gajadhar and Green [13] investigates the use of non-verbal textual cues and symbols in an online synchronous (chat) interaction. The focus is on examining the nature and the degree to which participants used nonverbal text-based communication to convey their messages. Their findings suggest that non-textual cues help convey the richness typically achieved in face-to-face communication and that these cues provide an important source of information to fully understand computer-mediated communication. Although this study did not examine dominance directly, it offers important insights for the study of non-verbal computer-mediated communication behavior.

The third study by Ocker [26] examined dominance in asynchronous groups by qualitatively analyzing the content of group communication. According to her study, "dominance is evidenced when a member has undue influence over the team's processes or work product." Coded dominance indicators were the following: attempt to control process or content, attempt to take ownership and responsibility of key concept development, attempt to claim of superiority, and comparatively higher volume of contribution. The findings of the study indicate that teams experiencing dominance demonstrate inhibited creativity because a dominant team member eschews compromise and consensus such that control and compliance are fostered.

Taken together, these studies offer important insights on how dominance has been and can be further studied in computer-mediated group communication.

## 2.2. Measurement of Dominance

Dominance constructs have been articulated in a variety of ways [6]. Widely used measures include scales to measure perceptions of dominance, as well as objective indicators. The latter is the focus of this study, as we are interested in the use of linguistic expression to identify particular dominance-related behaviors.

Liska [19] [20] has clearly articulated the dimensions of verbal and nonverbal dominance. Her studies identify specific patterns of communication behavior as being expressions of dominance. Norton [24] [25] analyzed communication styles extensively and developed 'Communicator Style Measures' which point to dominance behavior being expressed with identifiable characteristics. More specifically, above average number of comments, occupation of floor space (or air time in verbal discussions), use of various communication tactics, persuasion and influence methods have been treated as legitimate potential expressions of dominance.

Building on the concept of monopolization of air-time used in face-to-face discussions, dominance measures have been developed on the basis of total number of words, total number of statements, use of differentiation, use of emphasis and verbal cues used for automated dominance identification in meetings [31][28][27]. Communication measures such as number of comments, length of comments (number of words) and choice of words (e.g. number of self-references using "I"), are easily observable aspects of dominance in electronic communication.

In addition to the explicit use of words, dominance has been also been modeled on the basis of nonverbal activity indicators [17]. For instance, Gajadhar and Green [13] study the use of non-verbal textual cues and symbols in an online chat interaction. They focus on examining the nature and degree to which participants used text-based communication to convey their messages. To this end, they used a creative approach to analyze the chat content by focusing on textual emphasis cues, letters and sequence in words (i.e. orthographics), discourse markers, capitalization, formation of words with sounds (i.e. onomatopoeia), and abbreviations. The use of these cues helps participants convey their emotions and thoughts. For example, a set of upper case letters in a sequence demonstrated emphasis and an attempt to insist upon the message communicated. In addition, repetitive exclamation marks and similar cues serve as an expression of confidence and are used to communicate force, thus lending itself to dominance behavior. The study of dominance in electronic group interaction – though the analysis of explicit textual expressions, as well as implicit non-verbal cues –is poised to enhance our understanding of the dominance construct.

## 3. Research Methodology

To investigate dominance, we use manual and automatic content analysis. In manual content analysis, human coders identify instances of specific textual or non-textual cues in written communication. In automatic content analysis (also called automatic text analysis), computer programs automatically scan the content of textual documents to extract relevant information and generate descriptive summaries. Automatic text analysis methods have been applied to a variety of source documents both offline and online, particularly to web pages to analyze their content [34], or to classify them in pre-determined categories [35].

For this study, we selected several transcripts of electronic communications of groups solving a decision-making task. The text of the transcripts was analyzed to identify variables related to dominance. When applied to electronic group communication, textual analysis reveals important insights that are not typically available through other methods such as retrospective interviews or post-test questionnaires. For example, content analysis of group discussions can offer precise descriptions of the nature of group coordination and the pace of the interaction [1].

Consistent with our research objectives, the transcripts were parsed using automatic and manual methods to identify variables related to dominance. The comparison of algorithmic text mining with human coding techniques to analyze online communication has been used in other contexts before (e.g. Kontostathis et al. [18]. However, to the best of our knowledge, this is the first attempt to use this comparison to showcase the potential of automatic text analysis methods in virtual team communication.

### 3.1. Participants and procedures

Seven group transcripts were selected for the analysis. There were six members in each group. Participants were recruited from the student population at a large urban college in the Northeast of the United States. The teams were tasked with correctly identifying the guilty suspect in a fictitious cybercrime. Subjects remained anonymous in the course of the experiment, and received monetary compensation (flat fee plus variable amount depending on their individual and group performance).

The decision-making task was implemented as a "hidden profile task" [33][32]. Hidden profile tasks have been employed in the groupware literature to examine how information flows in group discussions and how final decisions are made [7][21].The task was a custom-developed case that presented three possible suspects (Alex Mansi, Mr. Nali and John Donahue) along with a set of clues in a computer security breach incident. Consistent with the guidelines of hidden profile tasks, the clues for the case were asymmetrically distributed among group members such that no member had the complete set of information required to solve the case.

At the beginning of the experimental session, each participant was given the text of the case with a partial set of clues. Before the group meeting, each member recorded his/her individual solution to the case (individual selection of guilty suspect). Then, within each group, members communicated electronically to discuss and solve the case together.

## 4. Data Analysis

Seven groups of six members each were randomly selected for the analysis. The participants were students enrolled in a business school in the Northeast of the US. The demographic description of the sample is 40% female, and an average age of 23.9 years. The transcripts of the selected group sessions were downloaded from the communication system and saved onto text files. Each transcript contained the comments that were exchanged along with a unique participant identifier and a time stamp. Table 1 shows the statistics for the number of comments, length of comments and number of words for all the seven groups, the minimum and maximum values and averages across the selected groups.

**Table 1. Sample Descriptive Statistics**

| Data Item | Total | Average (St. Dev) | Min | Max |
|---|---|---|---|---|
| Comments | 1,283 | 183.29 (94.9) | 101 | 388 |
| Length | 34,533 | 4933.29 (1,291.6) | 2,532 | 6,355 |
| Words | 7,255 | 1,036.43 (285.1) | 544 | 1,441 |

Automatic coding of comments (and "comment threads") was carried with a custom-developed parsing program that identified the variables outlined in Table 2. These variables have been previously used in dominance studies. Consistent with Ocker [26] we coded volume of contribution with comment count, comment length, and word count), and attempts to control the pace of the interaction (time references in our case). Other indicators of discourse such as question marks, exclamation points and self-references consistent with the literature ([13][17]) were also identified. A variable counting the references to the suspects' names (in the decision-making case) was also added.

Manual coding of chat transcripts was done according to Bales' [1] interaction process categories. The aim of the manual coding was to obtain a complete portrayal of the group interaction. In order to do so, we combined general descriptors of the communication along with specific indicators related to the context of our study.

**Table 2. Variables collected with automated teqniques**

| Numeric Variables | Description / Example |
|---|---|
| Comment Count | Count of each subject comment |
| Comment Length | Sum of character length of each subject's comments |
| Word Count | Sum of count of words of each subject's comments |
| All Caps Words | Count of words in all caps: "IT HAS TO BE ALEX" = 5 |
| Time References | Count of comments referring to time: "only 10 mins left" = 1 |
| Exclamation Points | Count of exclamation points: "good point #5!!!" = 3 |
| Question Marks | Count of question marks: "why Nali ??" = 2 |
| Self -References | Count of self -references: "I don't know" = 1 |
| Choice Reference | Count of references of choice: "so it is John or Nali???" = 2 |

Manually coded variables can be classified into two categories: task-related and socio-emotional. The socio-emotional category includes humor expression and humor appreciation and profanity. The task-related category encompasses case related questions and answers, and statement of preferred choice (directly and with pronouns). This category also includes process-related variables such as calling for a vote, trying to organize the discussion, attempting to get the discussion back on the task and hand, and indication of asymmetric information distribution.

Given the experimental manipulation implemented in this study, the realization of asymmetric information is an important "discovery" that when shared within a group, forces members to identify their unique pieces of information. Another variable manually coded was the mention of the suspects' directly using their names and indirectly, with the use of pronouns. Thus the variable *ChoiceReferencePro* is parallel to the one collected automatically. Table 3 presents a list of the manually coded variables along with their definitions.

**Table 3 Variables collected with manual techniques**

| Categorical Variables | Description / Example |
|---|---|
| Humor | An attempt at humor: "…it was the gardener in the kitchen with the candle stick" |
| Humor Appreciated | Appreciation for the aforementioned humor: "lol" or "ha ha" |
| Profanity | Comment containing profanity |
| Questions | Comments that ask a question (may or may not have question mark): "but why u say alex" |
| Answers | Comment that directly answers a previously asked question: "enough time for him to do it" |
| Call for Vote | Calls to vote on a final choice: "anyone who think its john type yes" |
| Organizational | Comments to attempt to organize discussion: "now lets analyze john" |
| Choice Reference Pro | Count of references of choice including pronouns: "I think it is John. Only he could have done it" = 2 |
| Refocus | Comments intended to refocus the group after side comments: "get back to work" |
| Asymmetric Information | Comments indicating information is not uniformly distributed to all group members |

The dependent variable in this study is a binary variable indicating Expression of Dominance (ED) at the comment level. Following the guidelines in Burgoon and Hale [5], a manual independent coding of dominance was performed by two of the authors working independently. To determine this categorical indicator, comments were coded based on an assessment of explicit and implicit textual cues indicative of dominance. More specifically, the comments were evaluated in terms of content (for example, repeated attempts to assert control over others by mentioning constantly a decision outcome), and in terms of tone (e.g. use of caps or exclamation signs). Inter-coder reliability calculated as the percentage of agreement is 0.90 (Cohen's Kappa = 0.75). The discrepancies were reconciled by reaching consensus on the most appropriate value for the ED indicator.

## 5. Results

Before proceeding to the analysis at the comment level, we examined the ED indicator at the individual and group level. To this end, we calculated the percentage of each group members' ED comments within their group and selected those individuals whose dominance percentage is larger than one standard deviation from the mean. Since there are 6 members in each group, we expect the mean percentage of dominance comments to be 16.7% in a group. In our sample, we find the mean percentage dominance comment is indeed 16.7% with standard deviation of 12.2%.

Using the threshold of dominance percentage larger than one standard deviation from the mean, we find that 8 participants from a total of 42 (7 groups with 6 members each), clearly exhibit dominance as shown in Figure 1. There was only one group where no dominant member emerged. Two groups had two dominant members and the rest had only one.

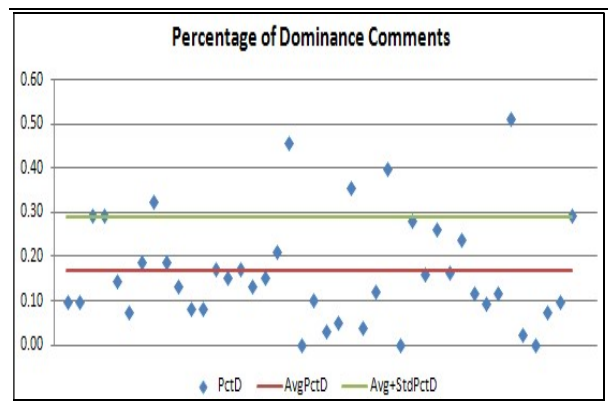

**Figure 1 Percentage of dominance comments with average + 1 standard deviation threshold**

The rest of our analyses are performed at the comment level. Given the expression of dominance indicator variable (ED) and the automatic and manually coded independent variables outlined above, we form three logit regression models to explore the relationship between chat comment features and expression of dominance, using the PROC LOGISTIC function in SAS.

**Table 4. Logit Regression coefficients for Model 1 (with automatically coded variables)**

| Parameter | Est. | Std. Error | Chi-Square | |
|---|---|---|---|---|
| Intercept | -1.20 | 0.51 | 5.65 | * |
| CommentLengthChar | -0.001 | 0.02 | 0.00 | |
| WordCount | -0.04 | 0.11 | 0.15 | |
| AverageWordLength | -0.20 | 0.12 | 2.96 | |
| Choice Reference | 2.43 | 0.17 | 218.35 | *** |
| AllCaps_Words | 0.06 | 0.06 | 0.78 | |
| TimeReferences | 5.57 | 1.49 | 14.06 | *** |
| CountExclamationPoi | 0.16 | 0.11 | 2.05 | |
| CountQuestionMarks | -0.92 | 0.30 | 9.57 | ** |
| SelfReferences | 0.27 | 0.18 | 2.10 | |
| Residual Dev 1102 | | | | |
| AIC: 1122 | | | | |

The selection of logistic regression was dictated by the binary nature of the dependent variable (ED). In logit models, two parameters provide goodness of fit information: the Akaike's Information Criterion (AIC) and the residual deviance. In both cases, smaller values are indicators of better fit, and several alternative models are typically compared.

Model 1 employs all of the automatically coded variables (from Table 2). The residual deviance is 1102.00 and AIC is 1395.86. The coefficients for each variable are shown in Table 4 (Significance is noted at the *0.05, **0.001 and ***0.0001 level). In addition to the intercept, the variables with significant coefficients are ChoiceReference (mentions of the solutions), time references and counts of question marks. In terms of fit, the AIC of the model is 1122 and the residual deviance is 1102. The AIC of model 1 is slightly lower than the one obtained for model 2, indicating that the automatically coded variables provide a better fit for the dominance indicator.

Model 2 employs all of the manually coded variables (from Table 3) and produces the results shown in Table 5. In this model, four variables have significant coefficients: questions, calls for votes, organizations and contains a mention of the solution (ChoiceReferencePro). The AIC of this model is 1192.5 and the residual deviance is 1170.5.

For comparison purposes, all the variables automatically detected and manually were used in Model 3. The results are shown in Table 6. In this case, the intercept is no longer significant and additional variables show significant coefficients. For this combined model, the AIC is lower than in the other two models, suggesting a better fit.

**Table 5. Logit regression coefficients for Model 2 (with manually coded variables)**

| Parameter | Est. | Std. Error | Chi-Square | |
|---|---|---|---|---|
| Intercept | -2.26 | 0.14 | 268.10 | *** |
| Humor | -0.62 | 0.96 | 0.42 | |
| HumorAppreciated | -0.14 | 1.62 | 0.00 | |
| Profanity | -0.61 | 1.05 | 0.34 | |
| Answers | -0.44 | 0.90 | 0.23 | |
| Questions | -1.49 | 0.32 | 21.04 | *** |
| CallForVote | 2.36 | 0.62 | 14.71 | *** |
| Organizational | 2.35 | 0.78 | 9.11 | ** |
| AsymetricInfo | 1.16 | 0.65 | 3.13 | |
| Refocus | -0.50 | 0.80 | 0.39 | |
| Choice ReferencePro | 2.15 | 0.16 | 176.40 | *** |
| Residual dev:1170.5 | | | | |
| AIC: 1192.50 | | | | |

**Table 6. Logit regression coefficients using Model 3 (with automatically and manually coded variables)**

| Parameter | Est | Std. Error | Chi-Square | |
|---|---|---|---|---|
| Intercept | -0.86 | 0.56 | 2.46 | |
| Humor | -0.27 | 0.97 | 0.08 | |
| HumorAppreciated | -0.38 | 1.64 | 0.05 | |
| Profanity | -0.50 | 1.05 | 0.23 | |
| Answers | -0.47 | 0.94 | 0.25 | |
| Questions | -0.994 | 0.39 | 6.51 | ** |
| CallForVote | 3.36 | 0.71 | 22.66 | *** |
| Organizational | 3.07 | 0.85 | 12.96 | ** |
| AsymetricInfo | 2.07 | 0.68 | 9.26 | ** |
| Refocus | -0.94 | 0.83 | 1.28 | |
| ContainsAnyPro | 1.88 | 0.31 | 37.19 | *** |
| CommentLengthChar | 0.02 | 0.02 | 0.88 | |
| WordCount | -0.17 | 0.12 | 2.14 | |
| AverageWordLength | -0.36 | 0.13 | 7.76 | ** |
| Contains_Any | 1.11 | 0.27 | 17.12 | *** |
| AllCaps_Words | 0.07 | 0.07 | 1.07 | |
| TimeReferences | 5.98 | 1.50 | 15.95 | *** |
| CountExclamationPoi | 0.29 | 0.13 | 5.66 | * |
| CountQuestionMarks | -0.76 | 0.37 | 4.26 | * |
| SelfReferences | 0.37 | 0.19 | 3.63 | |
| Residual dev: 1013.8 | | | | |
| AIC: 1053.8 | | | | |

The comparison of models shows that the set of automatically collected variables offers a reliable estimation of dominance, which is comparable to manually collected indicators. Although the combination of both yields a model with a better fit, the results of the two sets of variables (automatically detected and manually coded) are equivalent.

## 6. Discussion

The results of this study demonstrate the development and potential usefulness of identifying dominance using automatic textual analysis of chat transcripts. While some of the work was tailored to the topic of dominance, some of the other variables are applicable across a wide range of contexts. Therefore, generalized models based on textual analysis can be used to identify dominance and dominant individuals in computer-mediated group communication.

Traditional groupware research has promoted the idea that virtual interaction platforms generally create a level playing field, increase the quantity and quality of ideas and reduce the influence of dominant individuals. Our findings indicate that while this may be the case, even in situations where no single team member has complete information (due to the use of a hidden profile task), and group interaction occurs through an electronic medium that promotes equitable participation, dominance still emerges.

Since dominance is a likely occurrence in group interaction, its detection with automatic methods can offer real-time feedback of the nature of group discussions. Automatic detection of dominance offers the opportunity to apply interventions if necessary. We refrain from ascribing positive or negative connotations to the emergence of dominance in virtual teams. Depending upon the context, dominance can lead dysfunctional groups to produce an outcome, or it can deadlock functional groups. Regardless of the positive or negative connotations, the contribution of this study is to show that the detection of dominance within a group can be automated.

These results are subject to the limitations posed by the experimental setting and the sample. Specific findings related to dominance may be dependent upon the context of group interaction and the nature of the task. In addition, only several randomly selected transcripts were examined. While the sample of comments is rather large (almost 1,300), some of these comments are related to the dynamics of the discussion in each group and to the personality and participation of each individual member within a group. These limitations constraint the extent to which our models can be generalized across different situations where dominance emerges in computer-mediated communication. Nevertheless, we believe that this study serves as a significant and original articulation towards creating automated models for identifying dominance (and dominant entities) in electronic interaction.

### 6.1. Implications

Our study provides important implications for researchers and practitioners concerned with understanding or monitoring electronic interaction. Although it might not possible to assess dominance as a personality trait in a wide range of individuals, it is possible to identify dominance in computer-mediated group interaction. Depending upon the objectives of researchers and practitioners, dominance may (or may not) be desirable. Therefore, the implementation of automatic content analysis methods, such as the ones described in this paper, could provide early detection methods and the opportunity to administer particular interventions.

From a methodological point of view, automatic content analysis methods are a new tool to advance research in computer-mediated group interaction and it could be used for a variety of purposes. Compared to the time and costs associated with manual coding, automatic content analysis is a low-cost solution. The effectiveness of this approach hinges on the ability to identify relevant textual cues (both implicit and explicit) that could be used as indicators of the phenomenon being investigated.

For the study of dominance in virtual teams, our findings have important implications. Researchers will have an opportunity to examine existing models and create a new stream of research by investigating in more depth the role of dominance in processes and outcomes of group interaction. Practitioners will have an opportunity to improve monitoring of ongoing group communication and introduce new features in the design of virtual platforms to enhance or mitigate dominance, depending upon their objectives.

There are several avenues to extend this study. On the one hand, more groups transcripts could be coded an included for further analysis. On the other hand, the automatically coded variables could be used to parse the chat transcripts of other small or large groups in different settings. In either case, there is the larger issue of whether dominance is conducive to better or worse outcomes. With the automatic identification of dominance and an assessment of group outcomes in a large sample of groups, this research question can be systematically investigated.

## 7. Conclusion

The objective of this study is to demonstrate the feasibility of automating the identification and detection of dominance in computer-mediated group interaction. After coding a set of variables through automatic and manual methods, and comparing alternative models, we find that a subset of automatically collected variables offers a reliable estimation of dominance. These results are encouraging for researchers and practitioners interested in applying automatic text analysis methods of group communication to investigate specific constructs.